\documentclass[aps,prb,twocolumn]{revtex4}
\usepackage{graphicx}
\usepackage{xcolor}
\usepackage{siunitx}
\usepackage{amsmath}
\usepackage{bm}
\renewcommand{\vec}[1]{\boldsymbol{#1}}
\def\paraH2{{\it p}-H$_2$}
\def\he4{$^4$He}
\def\Am2{\AA$^{-2}$}
\def\gapx{\lower 2pt \hbox{$\buildrel>\over{\scriptstyle{\sim}}$}}
\def\lapx{\lower 2pt \hbox{$\buildrel<\over{\scriptstyle{\sim}}$}}

\begin{document}

\title{Absence of superfluidity in a quasi one-dimensional \\ parahydrogen fluid adsorbed inside carbon nanotubes} 

\author{Adrian Del Maestro}
\affiliation{Department of Physics, University of Vermont, Burlington, Vermont, 05405, USA}
\author{Massimo Boninsegni}
\affiliation{Department of Physics, University of Alberta, Edmonton, Alberta, Canada T6G 2J1}

\date{\today}

\begin{abstract}
A recent claim of superfluid behaviour of parahydrogen adsorbed inside armchair carbon nanotubes [M. Rossi and F. Ancilotto, Phys. Rev. B \textbf{94}, 100502 (2016)] is disproven by means of first principle computer simulations. Conclusive numerical evidence shows that the low temperature equilibrium  thermodynamic phase of parahydrogen adsorbed inside a $(10,10)$ armchair nanotube features a crystalline shell adsorbed on the inner surface of the tube, and a well separated quasi one-dimensional central column showing no evidence of possible fluid or superfluid behaviour.   Rather, the system is quasi-crystalline and {\em non-superfluid}; its physical character is qualitatively identical to that observed in similar confined settings extensively studied in the past. The incorrect prediction of superfluidity stems from the erroneous use of an effective one-dimensional interaction potential that does not describe parahydrogen at the center of the nanotube.
\end{abstract}
\maketitle

\section{Introduction}
A wealth of theoretical evidence, mostly accumulated over the past decade, strongly suggests that the putative superfluid phase (SFP) of  parahydrogen (\paraH2), predicted over four decades ago,\cite{ginzburg72} is unlikely to occur in practice. The problem with the original prediction  is its failure to take into proper account the depth of the attractive well of the interaction potential between two \paraH2 molecules, which imparts to the system a strong tendency to crystallize, even in reduced dimensions.\cite{boninsegni04,boninsegni13} If it occurs at all, superfluidity in \paraH2 may only be observable in small  clusters (thirty molecules or less), \cite{sindzingre,fabio,fabio2} which remain liquid-like at low temperature due to their relatively large surface to volume ratio. 
\\ \indent
Various theoretical and experimental avenues  have been
explored, aimed at suppressing crystallization in bulk \paraH2, thereby paving the way to a possible superfluid transition; these have  mainly attempted to exploit the (combined) effect(s) of disorder,  confinement, as well as the presence of external periodic potentials incommensurate with the preferred crystalline arrangement of \paraH2 molecules in their equilibrium solid phase -- all of these conditions are expected (and in some cases have been shown)  to expand the region of stability of fluid phases, with respect to the bulk.
\\ \indent
However, although confinement and disorder are known to lead to novel phases of matter,\cite{stan,dang}  and indeed there are theoretical predictions of enhanced superfluid response of \paraH2 clusters in nanoscale size confinement,\cite{omiyinka} none of the above-mentioned attempts has so far yielded any evidence of fluid-like behaviour of \paraH2, much less superfluidity at low temperature ($T$). Worth mentioning are, for instance, experimental studies of \paraH2 in the confines of porous vycor,\cite{bretz81,schindler96} all showing that the system is in a crystalline phase at low $T$, albeit with a different  structure\cite{azuah} than that of bulk \paraH2.
\\ \indent
On the theoretical side, there is strong evidence that merely embedding a fluid of \paraH2 molecules in  a foreign crystalline matrix, incommensurate with the equilibrium crystal structure of pure \paraH2,  does {\em not} result in a SFP.\cite{njp,turnbull,me16} It really seems unlikely that one may arrive at \paraH2 superfluidity in this way, as a low-density commensurate crystal is the only additional phase that can result from the presence of an external periodic potential. The same qualitative conclusion is reached even in the presence of a random distribution of scatterers.\cite{turnbull}
\\ \indent
A different kind of confinement, which has been long deemed favorable to the stabilization of novel phases of matter, consists of restricting particle motion to 
(quasi) one dimension (1D). This has motivated theoretical studies of quantum fluids such as $^4$He\cite{miller,moroni0} and \paraH2\cite{boninsegni13} in strictly 1D,
as well as in models of confinement aimed at describing the physical environment experienced by atoms and molecules moving inside a single carbon nanotube,\cite{gordillo} or the interstitial channel of a bundle of nanotubes,\cite{crespi} or nanopores.\cite{delma} 
\\ \indent
The low $T$ equilibrium properties of \paraH2 confined in long, narrow (around \SI{1}{\nano\meter} diameter) cylindrical channels, which constitute a reasonably realistic model of a
fluid of  \paraH2 molecules, e.g., in the pores of glass like vycor, or even a carbon nanotube,\cite{stan} have been recently studied systematically by first principle Quantum Monte Carlo (QMC) simulations.\cite{omi2} The main results of that study are that, on varying the attractive strength of the wall 
of the cylindrical pore, as well as its diameter, the equilibrium phase evolves from a single quasi-one-dimensional (1D) channel along the axis, to a cylindrical crystalline shell, to the simultaneous presence of a crystalline shell coating the inner surface of the cylinder and a (physically well separated) quasi-1D axial column. 
\\ \indent
All quasi-1D systems (i.e., the central column)  retain a strong propensity to crystallization, albeit somewhat reduced compared to the purely 1D system.  This is in stark contrast to what is observed in confined $^4$He \cite{delma} (due to its weaker inter-particle interactions). The results of Ref. [\onlinecite{omi2}]
yield no evidence of a topologically protected superfluid phase, as described by the Tomonaga-Luttinger Liquid Theory (TLL). \cite{tomonaga,luttinger,mattis,haldane}
\\ \indent 
Almost concurrently, however, the claim was put forth [M. Rossi and F. Ancilotto, Phys. Rev. B \textbf{94}, 100502 (2016), hereinafter referred to as RA] of a 
quasi-1D superfluid phase of \paraH2, in a physical setting very similar to that considered in Ref. [\onlinecite{omi2}],  ostensibly based on comparable calculations.   Specifically, it is contended that  inside a $(10,10)$ armchair carbon nanotube the equilibrium phase of an imbibed \paraH2 fluid should consist of a (solid) cylindrical shell adsorbed on the surface of the tube, as well as an inner quasi-1D channel, which is quite consistent with the findings of Ref.~[\onlinecite{omi2}]. However, 
it is concluded in 
RA  that the quasi-1D axial \paraH2 system displays superfluid character, in the TLL sense. \cite{RA}
\\ \indent
A key aspect  is that in RA the full three dimensional (3D) many-body Hamiltonian, which includes interactions between \paraH2 molecules as well as between \paraH2 and the carbon atoms, 
is replaced with a purely 1D one, only aiming at describing \paraH2 molecules along the axis. Such a modified Hamiltonian makes use of an effective renormalized pair-wise interaction  among \paraH2 molecules,
given by:
\begin{equation}
V_{\text{1D}}(z) = \frac{1}{\rho_{L}^2} \int d^2 r \int d^2 r' V(\vec{r}
            -\vec{r}')\rho(r) \rho(r')
\label{eq:V1D}
\end{equation}
where  $\vec{r} = (r,\varphi,z)$ is a vector in cylindrical coordinates, 
$\rho(r)$ is the radial particle number density, $\rho_{L} = N/L$ is the linear
density for $N$ total \paraH2 molecules inside a carbon nanotube of length $L$, and $V$ is the bare \paraH2 intermolecular pair potential.
It is important to note that, in RA, there is no cut-off in the radial integrations in (\ref{eq:V1D}), i.e., they extend
from the axis of the tube all the way to its surface, thereby including {\em all} the \paraH2 molecules in the system.
\\ \indent
RA assert that the presence of the \paraH2 shell adsorbed on the inner surface of the  tube has the effect of dramatically weakening the effective pair-wise interaction among molecules in the quasi-1D  central column. Indeed, the effective pair potential $V_{\rm 1D}$ arising from  (\ref{eq:V1D})   is strongly renormalized with respect to the
bare pair potential. In particular, the depth of the attractive well of $V_{\rm 1D}$ is  about three times  shallower than that of a typical, accepted \paraH2 
intermolecular potential, e.g., the  Silvera-Goldman.~\cite{SG} Such an effective interaction is much more similar to that  between two \he4 atoms,\cite{aziz} and is 
therefore scarcely surprising that the 1D simulations carried out in RA, based on (\ref{eq:V1D}),  find a strong propensity towards superfluidity (again in the sense of TLL theory), since this is consistent with 1D $^4$He physics.\cite{moroni0}

In this paper, we show that (I) $V_{\rm 1D}$ does not provide a
realistic (much less \emph{ab initio}) model of intermolecular interactions in the central 1D column of \paraH2 confined inside $(10,10)$ nanotubes and 
(II)  when performing an exact simulation of the full 3D microscopic system introduced by RA, we observe no evidence of superfluidity in the central column, but instead find a quasi-crystalline phase, undergoing spinodal decomposition below the 1D equibrium density of \paraH2, namely, $\sim\SI{0.22}{\angstrom^{-1}}$.  In other words, the physics of the axial \paraH2 system rather closely reproduces the theoretical 1D limit.
Thus, contrary to what stated in RA, the confines of a carbon nanotube prove no more promising an environment for the stabilization of a superfluid phase of \paraH2 than any of the ones that have been considered in the past.
\\ \indent
The remainder of this paper is organized as follows: in Sec.~\ref{mc} we introduce a microscopic model of \paraH2 confined inside a $(10,10)$ armchair carbon nanotube and provide computational details; in  Sec.~\ref{sere} we illustrate our results and offer a theoretical interpretation. Finally, we outline our conclusions in Sec. ~\ref{conc}.  

\section{Model and Calculation}\label {mc}
We considered a collection of $N$ point-like particles (\paraH2 molecules), moving in  three dimensions in the presence of an external potential arising from a lattice of $2000$ static, identical carbon atoms, aligned
to form  a section of length $L = \SI{123.08}{\angstrom}$ of a $(10,10)$ armchair nanotube of radius \SI{6.793}{\angstrom} (we neglect curvature as in RA).\cite{note}
Periodic boundary conditions are utilized in the axial direction, whereas in the other two they are immaterial as the system is confined.  
The quantum-mechanical many-body Hamiltonian is: 
\begin{equation}\label{ham}
\hat {\cal H} = -\lambda\sum_{i=1}^N\nabla_i^2 +\sum_{i<j}V(r_{ij})+\sum_{i\sigma}U(|{\bf r}_i-{\bf R}_\sigma|)
\end{equation}
Here, $\lambda=12.031$ K\AA$^2$, $V$ is the interaction potential between any two \paraH2 molecules, \cite{SG} only depending on their relative distance $r_{ij}\equiv |{\bf r}_i-{\bf r}_j|$, whereas $U$ represents the interaction of each \paraH2 molecule with a C atom.\cite{pillo} The above  Hamiltonian is identical to that reported in RA.  It should be noted that refinements of $V$ have been recently proposed, which afford a more accurate reproduction of the experimental equation of state of \paraH2 at high pressure,\cite{omi3} and that such potentials were used in the most recent studies of \paraH2 in confinement.\cite{omiyinka,omi2} 
\\ \indent
We studied the low temperature physical properties of Eq.~(\ref{ham}) by means of first principle computer simulations based on the Worm
Algorithm in the continuous-space path integral representation.\cite{worm,worm2} 
The most important aspects of this well-established method to be emphasized here, are that it enables one to compute numerically exact thermodynamic properties of Bose systems at finite temperature, directly from the microscopic Hamiltonian, in particular  energetic, structural and superfluid properties. 
\\ \indent
Technical details of the simulation are standard, and we refer the interested reader to Ref. [\onlinecite{worm2}]. We adopted the usual high-temperature approximation for the many-particle propagator accurate up to order $\tau^4$, and all of the results reported here are extrapolated to the $\tau\to 0$ limit;  in general, we found that a value of the imaginary time step 
$\tau=1/320$ K$^{-1}$ yields estimates that are indistinguishable from the extrapolated ones, within the statistical errors of the calculation. We obtained results in the temperature range $1$ K $\le T \le 4 $ K and observed  no discernible variation of the physical behaviour. In particular, within this range of temperature, even the estimates of the energy remain unchanged within our statistical uncertainties ($\sim$ 0.01\% on the total energy per \paraH2 molecule). Thus, the structural results presented here can be regarded as representative of the ground state of the system and can be directly compared with the $T=0$ results of RA.
\\ \indent
We point out that this study does not pose any particular challenges,  as exchanges of indistinguishable particles are virtually absent in this system.  The same results could in principle be obtained by means of a straightforward implementation of the path integral Monte Carlo methodology.\cite{jltp}

\section{Results}\label{sere}
We find, in agreement with the  simulations of RA of model (\ref{ham}), that  the \paraH2 adsorbate first forms a crystalline cylindrical shell of radius $\sim 3.7$ \AA. As the two-dimensional (2D) density of this surface layer exceeds $\sim 0.107$ \AA$^{-2}$,  molecules begin to occupy the central part of the tube, lining up along the axis.
\begin{figure}[h]
\centerline{\includegraphics[height=2.4in]{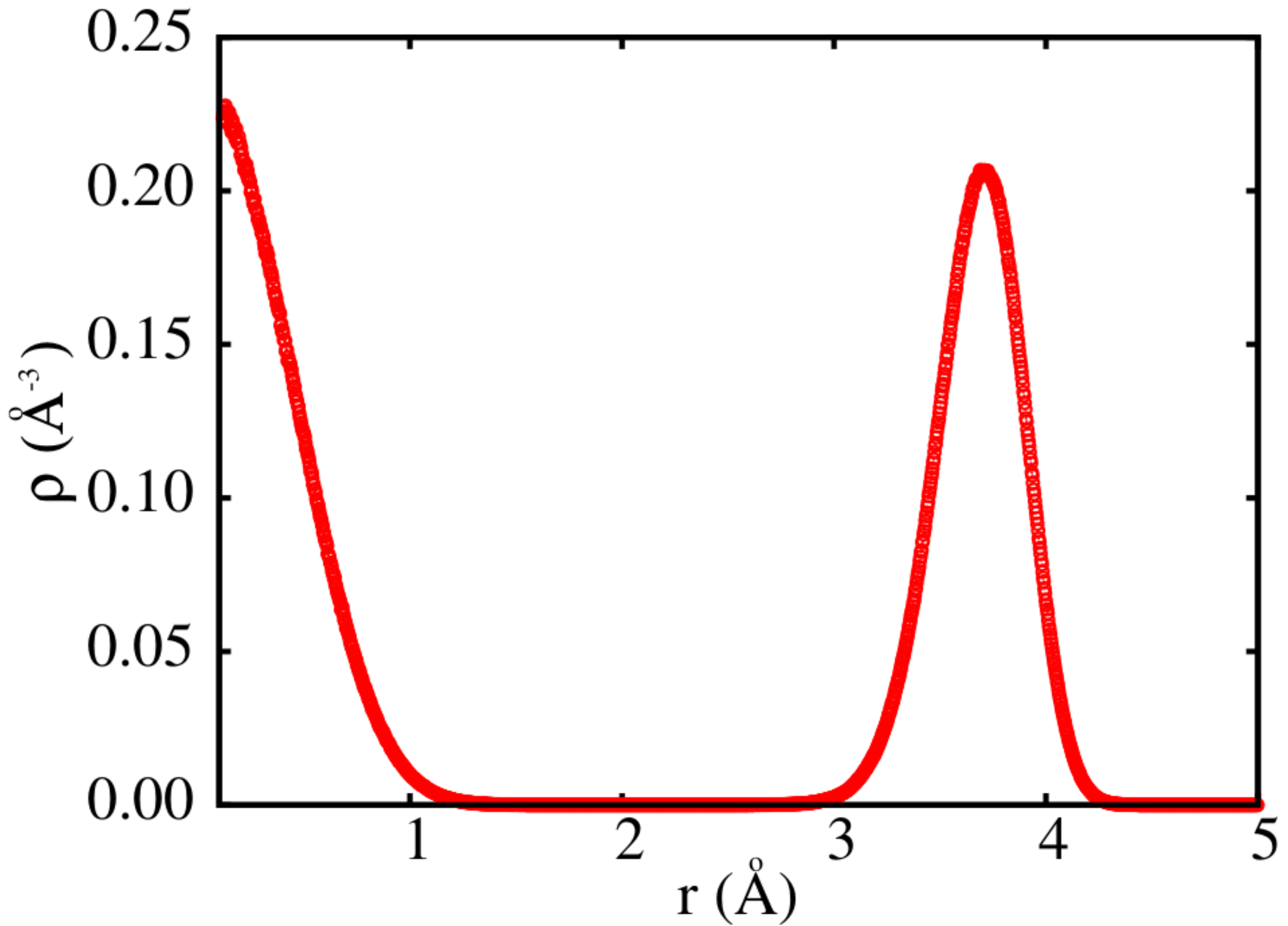}}
\caption{{\em Color online}. 
Radial density profile of \paraH2 inside a $(10,10)$ armchair nanotube, modeled as explained in the text. The axis of the tube is at $r=0$. The 2D
density of the layer of molecules adsorbed on the inner surface of the tube, of radius $\sim 3.7$ \AA, is  close to 0.107 \AA$^{-2}$. The 1D linear density of the inner column is 0.252 \AA$^{-1}$. There are in total $N=337$ molecules in this simulated system.}
\label{f1}
\end{figure}

Fig.~\ref{f1} shows computed (cylindrically averaged) radial density profiles of \paraH2 inside the nanotube. The outer shell of radius 3.7 \AA\ and the inner column are separated by an essentially empty region, almost 2 \AA\ wide. The 1D linear density of the inner column is 0.252 \AA$^{-1}$. As mentioned above, this density profile remains the same in the temperature range explored here; its main features are ({\em a}) no overlap between the central column and the outer shell ({\em b}) the width of the inner column, arising from zero-point motion alone, is of the order of half an \AA\ ({\em c}) no exchanges of indistinguishable \paraH2 molecules occur at low $T$, not between the central column and the shell, 
 within she shell, which is solid, nor within the column. All of these observations are consistent with the full simulations performed in RA.
\\ \indent
We observe sharper density features, with narrower peaks than those previously reported for \paraH2 confined inside cylindrical pores of similar radii in weakly attractive substrates,\cite{omi2} as a result of the increased \paraH2 density within the confines of a carbon nanotube, a strong adsorber.  The  enhanced molecular localization, and in particular the lack of any observed particle density between the central quasi-1D column and surrounding shell, and the ensuing absence of quantum exchanges, are {\em not} signatures of an energetically favorable environment for superfluidity.\cite{kul} The lack of overlap between the wave function of molecules near the axis and those in the shell,
renders these two systems {\em de facto} distinguishable, precluding the use of $V_{\rm 1D}$ as given by (\ref{eq:V1D}), i.e., without any cut-off beyond the
axial core for the radial integrations. 
\\ \indent
According to the TLL theory, the (quasi)superfluid character of a quasi-1D system can be inferred from the long-distance behaviour of the pair correlation function
$g(r)$, which oscillates around unity, the amplitude of the oscillations decaying as $r^{2/K}$ as $T\to 0$. The value of the Luttinger parameter $K$ determines whether the system displays quasi-superfluid or quasi-crystalline behaviour. Specifically, if $K>2$ the static structure factor will develop (Bragg) peaks at reciprocal lattice vectors, which is the experimental signature of a crystalline solid. On the other hand, if $K<1/2$ the system features a robust propensity to superflow, stable against pinning by a periodic potential whereas for $1/2<K<2/3$, the system is unstable to pinning but stable against localization due to weak disorder. \cite{giamarchi}  Thus, a robust prediction of superfluidity coming from a theoretical calculation such as that performed here (or in RA) can be based on a value of $K$ significantly less than 2.
\\ \indent
 We can investigate the value of $K$ for the central column of the nanotube by directly measuring the pair correlation function via quantum Monte Carlo.
\begin{figure}[th]
\centerline{\includegraphics[height=2.6 in]{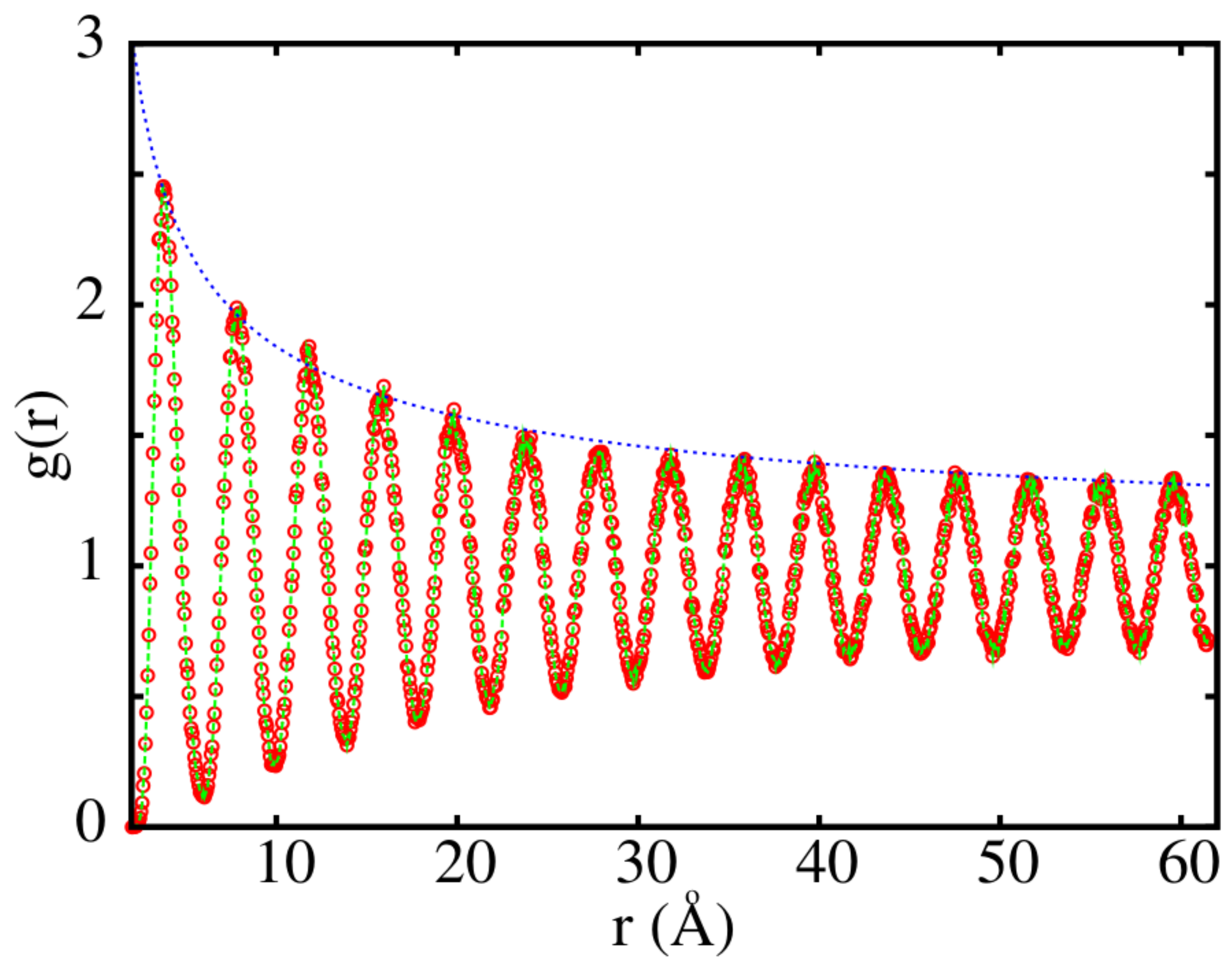}}
\caption{{\em Color online}. Reduced 1D pair correlation function of \paraH2 molecules along the axis of the carbon nanotube, computed  at $T$=4 K and in correspondence of a 
linear 1D density of 0.252 \AA$^{-1}$. Dotted line is a fit to the maxima of the function with the expression $f(r)=1+Ar^{-2/K}$.}
\label{f2}
\end{figure}
Fig. \ref{f2} shows the resulting value of $g(r)$ along the axis of the nanotube, in correspondence of the same 1D linear density of Fig. \ref{f1}, namely 0.252 \AA$^{-1}$, computed at  $T$=4 K. Upon fitting the maxima of the function to the expression $f(r)=1+Ar^{-2/K}$ we find that $K \approx 4$, a 20\% reduction compared to the 1D value of $K \approx 5$ estimated in this work by carrying out a purely 1D calculation of a system of the same size and density and temperature. 
\\ \indent
The physical mechanism responsible for the reduction in $K$ is identical to that observed in Ref. \onlinecite{omi2} for \paraH2 inside cylindrical channels in weak substrates.   It arises from the effective softening of the repulsive core of the interaction between two \paraH2 molecules moving on the axis, as a result of transverse zero-point excursions away from it (the outer \paraH2 shell plays no role). However, even in the more weakly confined case, this effect is insufficient to drive $K$ significantly below a value of 2, where superfluid correlations would be expected. Moreover, Fig.~\ref{f1} shows that in the more strongly confined system considered here, zero-point transverse excursions in the central column are even smaller (as deviation away from the axis cost potential energy).  Thus, while a precise 
determination of $K$ would require performing a complete finite-size and temperature scaling analysis of $g(r)$, the results shown here allows us to conclude that the reduction in $K$ is insufficient to sustain quasi-superfluidity, and that the physics of the central column largely remains equivalent to that of a purely 1D system.
\\ \indent
It is known that 1D \paraH2 is a quasi-crystal at $T=0$, with a value of $K\sim 3.5$ at the equilibrium  density ($\rho_e\sim 0.22$ \AA$^{-1}$), growing monotonically as the system is compressed and remaining $> 2$ all the way down to the spinodal density ($\approx 0.21$ \AA$^{-1}$), below which the system breaks into ``puddles'' and the TLL theory no longer applies.\cite{boninsegni13} To illustrate even further how the physics of \paraH2 inside a carbon nanotube is essentially that of 1D \paraH2, we now asses the claim made by RA that thermodynamically stable phases of the quasi-1D \paraH2 fluid in the central column exist, of much lower density than in the 1D system. 
\begin{figure}[t]
\centerline{\includegraphics[height=3.6 in]{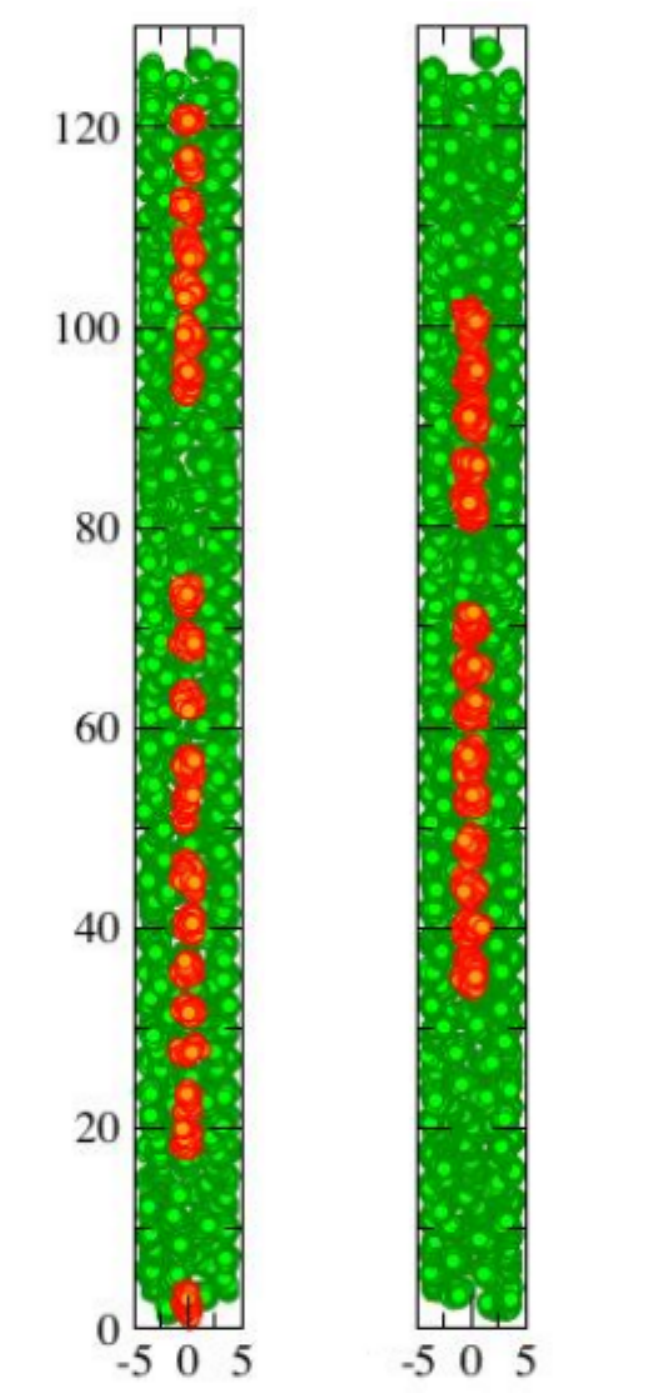}}
\caption{{\em Color online}. Snapshots of many-particle configurations (imaginary time world lines) for two different simulations of a \paraH2 fluid inside a 
carbon nanotube, modeled as in (\ref{ham}). Green beads refer to molecules in the outer shell, whereas red beads represents molecules along the axis of the tube. The effective 1D density is 0.163 \AA$^{-1}$ (left) and 0.114
\AA$^{-1}$ (right). The temperature of the simulations is 2 K.}
\label{f3}
\end{figure}
\\ \indent
Fig. \ref{f3} shows instantaneous configurational snapshots of the \paraH2 fluid adsorbed inside the nanotube, at a temperature $T=2$ K. Specifically, we show the discretized imaginary time ``world lines"\cite{jltp} of the various particles by means of ``beads" of different colors, namely green ones represent \paraH2 molecules in the solid layer coating the inner surface of the tube, whereas red ones refer to molecules along the central axis. In both cases, the 2D density of the layer coating the inner surface is 0.107 \AA$^{-2}$, whereas the nominal 1D  densities of the system in the inner column are respectively 0.163 \AA$^{-1}$ and 0.114 \AA$^{-1}$, i.e., both well below the 1D spinodal density of 0.21 \AA$^{-1}$. As clearly seen, in both cases the 1D system breaks down into ``puddles", whose linear density can be roughly visually estimated to be close to $\sim 0.22$ \AA$^{-1}$, i.e., the 1D equilibrium density.\cite{boninsegni13} Although Fig. \ref{f3} only shows
a single snapshot for the two densities, analogous and consistent observations of this structure were repeatedly made during the course of long simulation runs, 
for the two cases illustrated and in the entire temperature range considered  here. This is of course consistent with what one expects based on the results presented above.
\\ \indent
The conclusion is that the physics of the quasi-1D column of \paraH2 along the
central axis of the tube, closely mimics that of purely 1D \paraH2. In
particular, only a quasi-crystalline phase is observed, with no evidence of
possible superfluidity in the restricted 1D sense.

\section{Discussion}\label{conc}
Our first principles simulations of \paraH2 confined inside $(10,10)$ armchair carbon nanotubes show no evidence of a superfluid phase, in contrast to what was reported by RA using an effective one dimensional model of the same geometry.  The interior of a nanotube provides a strongly attractive environment for \paraH2 molecules, and only (quasi)crystalline phases are observed to occur at low $T$. We find an enhanced propensity toward crystallization as compared to the previously studied case of \paraH2 inside cylindrical pores.\cite{omi2}
\\ \indent
Our fully microscopic simulation results for the radial density of molecules inside the nanotube as shown in Fig.~\ref{f1} are in complete agreement with those presented by RA in their Fig.~1.  However, instead of  performing a simulation of Eq.~\eqref{ham}, as we did in this work, in RA the computed 
radial density profile is utilized as an input into Eq.~\eqref{eq:V1D} to obtain an effective strictly one dimensional interaction potential.  
All of the subsequent results for the Luttinger parameter $K$ reported in RA are based on this effective model, which is not proven to offer a quantitatively accurate  description of molecular structure or energetics inside the nanotube.  This is especially concerning, as interactions in their effective 1D model are \emph{strongly} renormalized, featuring a much shallower attractive potential well (about one third of that of the bare \paraH2 potential). The resulting potential is very similar to the bare interaction between \he4 atoms, but as the mass of bosonic \paraH2 molecules is half that of \he4, it is not at all surprising that  signatures of superfluidity
are found in RA. Additional claims of thermodynamically stable low density liquid phases are also in agreement with what is known to exist in 1D \he4.\cite{moroni0}

\begin{figure}[h]
\centerline{\includegraphics[height=2.3 in]{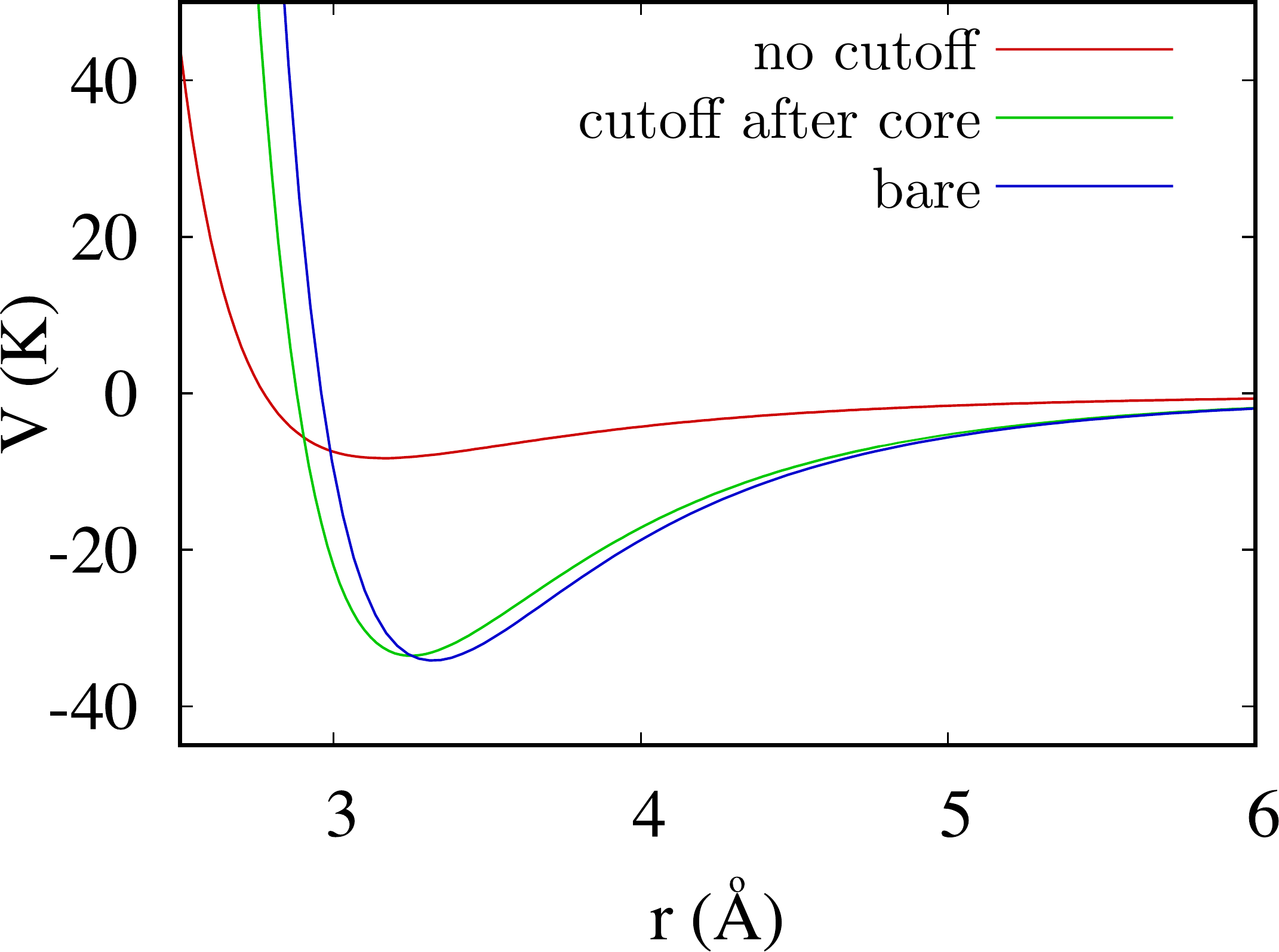}}
\caption{{\em Color online}. A comparison of the bare interaction potential as a function of separation between two \paraH2 molecules with the effective potential computed via Eq.~\eqref{eq:V1D} with a radial cutoff placed either between the central column and first density shell or past the outer shell. Using an unphysically large cutoff leads to a spurious softening of the potential.}
\label{f4}
\end{figure}

The effective potential utilized in by RA was previously discussed by one of us,\cite{adrian2} in the context of the study of superfluid $^4$He adsorbed inside cylindrical pores of nanometer size diameter. The considerably weaker \emph{bare} potential between \he4 atoms dramatically reduces the energy barrier for spatial fluctuations in the radial direction in that system and leads to enhanced atomic exchanges between the central column and surrounding cells.  This is confirmed by the radial density profiles, which always exhibit a finite density of particles between shells, or by measuring the spatially resolved superfluid density\cite{kul} which tracks the particle density. For \he4 inside nanopores, the wavefunction of atoms in the core penetrates far into the radial direction and $V_{\rm 1D}$ may provide some qualitative insights into the origin of the reduction in $K$ from its known purely $1D$ value.
\\ \indent
However, as seen in Fig.~\ref{f1}, \paraH2 inside carbon nanotubes exhibits strong localization and molecules near the axis do not participate in any exchanges (as evidenced by zero radial density between the column and first shell).  In these conditions, one may only utilize Eq.~\eqref{eq:V1D} by inserting a cutoff in the radial direction where the density vanishes ($\sim \SI{1.2}{\angstrom}$ in this case).  To phrase it differently, the delocalization of molecules on the axis arises exclusively from zero point motion, and the outer \paraH2 shell has no effect on the effective interaction between two molecules on the axis, contrary to what stated in RA.
We illustrate the effect of using this modified cutoff on the effective 1D potential in Fig.~\ref{f4}, along with unrestricted integration employed by RA and the bare interaction potential.\cite{notesimple}  The resulting effective potential differs only minimally from the bare potential, and a 1D simulation employing it will exhibit none of the superfluid traits observed in RA using the (spuriously) weaker one (line labelled ``no cutoff" in Fig. \ref{f4}).
\\ \indent
Summarizing, quantum Monte Carlo simulations of a realistic model of \paraH2 in the confines of an armchair (10,10) carbon nanotube of radius $\sim$ 6.7 \AA, yield no evidence of possible superfluid behaviour. On the contrary, they display behavior that is consistent with previous work in similar settings; the system has a strong propensity towards crystallization. These conclusions are at variance with a recent claim of a possible superfluid phase of \paraH2 in 
these physical settings due to an incorrect use of a effective one dimensional model that does not accurately capture the physical environment experienced by \paraH2 molecules along the axis of the nanotube.

\section*{Acknowledgements}

This work was supported  by the Natural Science and Engineering Research Council of Canada. Computing support of Westgrid is gratefully acknowledged.


\begin{thebibliography}{0}
\bibitem{ginzburg72}
V. L. Ginzburg and A. A. Sobyanin,  JETP Letters {\bf 15}, 242 (1972).
\bibitem{boninsegni04} M. Boninsegni, Phys. Rev. B {\bf 70}, 125405 (2004).
\bibitem{boninsegni13}
M. Boninsegni, Phys. Rev. Lett. {\bf 111}, 235303 (2013).
\bibitem{sindzingre} 
P. Sindzingre, D. M. Ceperley and M. L. Klein, 
Phys. Rev. Lett. {\bf 67}, 1871 (1991).
\bibitem{fabio}
F. Mezzacapo and M. Boninsegni, Phys. Rev. Lett. {\bf 97}, 045301 (2006).
\bibitem {fabio2}
F. Mezzacapo and M. Boninsegni, Phys. Rev. A {\bf 75}, 033201 (2007).
\bibitem{stan}
See, for instance, G. Stan, S. Gatica, M. Boninsegni, S. Curtarolo and M. W. Cole, Am. J. Phys. {\bf 67}, 1170 (1999), and references therein.
\bibitem{dang}
L. Dang, M. Boninsegni and L. Pollet, Phys. Rev. B  {\bf 79}, 214529 (2009).
\bibitem{omiyinka}
T. Omiyinka and M. Boninsegni, Phys. Rev. B {\bf 90}, 064511 (2014).
\bibitem{bretz81}
M. Bretz and A. L. Thomson,
Phys. Rev. B {\bf 24}, 467 (1981).

\bibitem{schindler96}
M. Schindler, A. Dertinger, Y. Kondo and F.  Pobell, 
Phys. Rev. B {\bf 53}, 11451 (1996).

\bibitem{azuah}
P. E. Sokol, R. T. Azuah, M. R. Gibbs and S. M. Bennington,
J. Low Temp. Phys. {\bf 103}, 23 (1996).

\bibitem{njp}
M. Boninsegni, New J. Phys. {\bf 7}, 78 (2005).
\bibitem{turnbull}
J. Turnbull and M. Boninsegni, Phys. Rev. B {\bf 78}, 144509 (2008).
\bibitem{me16}
M. Boninsegni, Phys. Rev. B {\bf 93}, 054507 (2016).
\bibitem{miller}
E. Krotscheck and M. D. Miller, Phys. Rev. B {\bf 60}, 13038 (1999).
\bibitem{moroni0}
M. Boninsegni and S. Moroni, J. Low Temp. Phys. {\bf 118}, 1 (2000).
\bibitem{gordillo}
M. C. Gordillo, J. Boronat and J. Casulleras, Phys. Rev. B {\bf 61}, R878 (2000).
\bibitem{crespi}
M. Boninsegni, S.-Y. Lee and V. H. Crespi, Phys. Rev. Lett.  {\bf 86}, 3360 (2001).


\bibitem{delma}
A. Del Maestro, M. Boninsegni and I. Affleck, Phys. Rev. Lett. {\bf 106}, 105303 (2011).
\bibitem{omi2}
T. Omiyinka and M. Boninsegni, Phys. Rev. B {\bf 93}, 104501 (2016).

\bibitem{tomonaga} S.-I. Tomonaga, Prog. Theor. Phys. \textbf{5}, 544 (1951).
\bibitem{luttinger} J.~M. Luttinger, J. Math. Phys. \textbf{4}, 1154 (1963).
\bibitem{mattis} D. C. Mattis and E. H. Lieb, J. Math. Phys. \textbf{6}, 304 (1965).
\bibitem{haldane} F. D. M. Haldane, Phys. Rev. Lett. \textbf{47}, 1840 (1981).
\bibitem{RA}
M. Rossi and F. Ancilotto, Phys. Rev. B {\bf 94}, 100502 (2016).
\bibitem{SG} I. Silvera and V. Goldman, J. Chem. Phys. {\bf 69}, 4209 (1978).

\bibitem{aziz} R. A. Aziz, V. P. S. Nain, J. S. Carley, W. L. Taylor, and G. T. McConville, J. Chem. Phys. \textbf{70}, 4330 (1979).

\bibitem{note}
The positions of the carbon atoms are obtained using TubGen online. \\ See, http://turin.nss.udel.edu/research/tubegenonline.html

\bibitem{omi3}
T. Omiyinka and M. Boninsegni, Phys. Rev. B {\bf 88}, 024112 (2013).
\bibitem{pillo}
D. Y. Sun, J. W. Liu, X. G. Gong, and Z-F. Liu, Phys. Rev. B {\bf 75}, 075424 (2007).
\bibitem{worm}
M. Boninsegni, N. Prokof'ev and B. Svistunov, Phys. Rev. Lett. {\bf 96}, 070601 (2006).
\bibitem{worm2}
M. Boninsegni, N. Prokof'ev and B. Svistunov, Phys. Rev. E {\bf 74}, 036701 (2006).
\bibitem{jltp}
See, for instance, M. Boninsegni, J. Low Temp. Phys. {\bf 141}, 27 (2005).

\bibitem{kul} B. Kulchytskyy, G. Gervais, and A. Del Maestro,  Phys. Rev. B \textbf{88}, 064512 (2013).

\bibitem{giamarchi} T. Giamarchi and H. Schulz, Phys. Rev. B \textbf{37}, 325 (1988).

\bibitem{adrian2}
A. Del Maestro, Int. J. Mod. Phys. B {\bf 26}, 1244002 (2012).

\bibitem{notesimple}
For simplicity, a Lennard-Jones  potential was utilized to obtain the curves shown in Fig. \ref{f4}, one with $\epsilon=34.16$ K and $\sigma=2.96$ \AA, i.e., the
standard values for \paraH2.

\end{thebibliography}
\end{document}